# Battery selection for optimal grid-outage resilient photovoltaic and battery systems

Stamatis Tsianikas[a], Jian Zhou[a], Nooshin Yousefi[a] and David W. Coit[a]

[a] Department of Industrial and Systems Engineering, Rutgers University, Piscataway, NJ 08854-8065, USA

## Abstract

The first and most important purpose of the current research work is to investigate the effects that different battery types have on the optimal configuration of photovoltaic (PV) and battery systems, from both economic and resilience perspectives. Many industry reports, as well as research papers, have already highlighted the crucial role that storage systems have in the coming years in the electricity sector, especially when combined with renewable energy systems (RES). Given the high cost of storage technologies, there is an urgent need for optimizing such integrated energy systems. In this paper, a simulation-based method is adopted and improved, in order to compare different battery types based on their characteristics. After the introduction of four different battery types, i.e. lead-acid, sodium sulphur, vanadium redox and lithium-ion, the mathematical model for the optimization problem is presented, along with the required explanations. Subsequently, a case study is described and the numerical assumptions are defined. Therein, our specific focus addresses the different values that the four battery types possess in three critical parameters, i.e. battery cost, efficiency and depth of discharge (*DoD*). Finally, results and discussion are provided in an illustrative and informative way. This model provides a useful guide for relevant future work in the area, and also serves as a baseline for more comprehensive methodologies regarding optimal sizing of photovoltaic and battery systems.

## 1. Introduction

In recent years, using photovoltaic energy generation to reduce the usage of fossil fuel and provide power resiliency has received considerable attention. In 2014, the International Energy Agency projected that, under its "high renewables" scenario, PV could supply 16% of global electricity generation by 2050 [1]. Combining solar PV energy system with energy storage can compensate for the intermittency nature of solar energy. Battery technology is one of the most popular energy storages currently used. However, battery price is still relatively expensive when compared with other energy storage technologies. As a result, optimization of the size combination of PV array and battery is becoming an important research challenge. Shrestha and Goel [2] proposed a method to select the optimal combination of stand-alone PV and battery unit to satisfy the load demand based on energy production simulation. Kaushika et al. [3] presented a linear programming model to optimize the combination of PV and battery storage to minimize the loss of power supply. Many researchers have investigated systems combining PV and battery storage using different procedures and approaches [4-6].

Energy storage technology is critical to the future development of renewable energy. Haghi et al. [7] analyzed the effects and cost efficiency of using energy storage in renewal energy systems. Miller et al. [8] discussed the improvement of power quality and reliability of power system using battery energy technology., Among all available storage technologies, battery technology is one of the most widely used storage devices for power system applications [9]. Some important battery properties vary between different types of batteries, including battery efficiency, battery price, life span, energy density and *DoD*. Of all current types of batteries, four of them are more suitable for power system applications and are compared in the proposed model. A brief description and details of these batteries are as follow:

**Lead-acid:** Each cell of a lead-acid battery comprises a positive electrode of lead dioxide and a negative electrode of sponge lead, separated by a micro-porous material and immersed in an aqueous sulfuric acid electrolyte [10]. Lead acid batteries are a preferred solution for renewal energy systems (RES) which has power rating of approximately 100 kW.

**Sodium sulphur (NaS):** Consists of molten sulfur at the positive electrode and molten sodium at the negative electrode separated by a solid beta alumina ceramic electrolyte [10]. This battery operates at temperatures of 300ºC, which makes it less applicable for renewal energy systems (RES) with long periods of time without surplus power from renewable sources.

**Vanadium redox**: uses hypovanadous/vanadous and vanadyl/vanadic ion exchange membrane separating two electrolyte solutions. There are no commercial products for small scale RES. However, in the long term there is a high potential for storage systems with a high ratio of energy content (kWh) to power rating (kW) [11].

**Lithium-ion (Li-ion)**: Lithiated metal oxide is used in the cathodes and graphitic carbon with a layer structure used for the anode. Having high energy density and high efficiency, this battery is suitable for many RES applications.

In this paper, an extension of [12] is proposed to find the optimal combination of PV and battery considering different types of battery storages. The modified model is defined and the presentation of the case study follows. Results are presented using a comparative figure and discussion.

## Nomenclature

| | |
|---|---|
| *LOLP* | Loss of load probability, % |
| *VOLL* | Value of Lost Load, $/kWh |
| $B_r$ | Battery capacity, kWh |
| *b* | Battery price, $/kWh |
| $B_{min}$ | Minimum battery energy value, kWh |
| *DoD* | Maximum depth of discharge of battery, % |
| *AEL(t)* | Amount of energy lost, kWh |
| $Q_B(t)$ | Energy stored in battery at time *t*, kWh |
| *P(t)* | Power generated by PV array at time *t*, kW |
| *D(t)* | Load demand at time *t*, kW |
| *d* | Proportion of demand that needs to be met |
| *e* | Efficiency of discharging/charging process of battery, % |
| *η* | Energy conversion efficiency of PV array, % |
| $I_t$ | Solar irradiation at time *t*, W/m$^2$ |
| *A* | PV array area, m$^2$ |
| *T* | Grid outage duration, hours |
| *β* | Constraint of *LOLP*, % |
| CAIDI | Customer Average Interruption Duration Index, hours/interruption |
| SAIFI | System Average Interruption Frequency Index, interruptions/year |

## 2. Methodology

The optimization problem that we are focused on is to minimize total system cost (*TSC*) under specific reliability constraints. It is assumed that PV array size has already been selected and the decision variable is the size of battery capacity that is going to be integrated with the PV system.

The optimization formulae model proposed in the following part are mainly based on [12]. However, the previous assumption made in the existing model, that battery is always fully charged at the beginning of each grid outage, is not consistent with the real cases. It is clear that the initial charging state of battery storage affects the amount of energy that can be satisfied during gird interruptions, which needs to be taken into account for optimizing battery size. Therefore, we assume that battery charging state when grid outage occurs follows a continuous uniform distribution between 0 and 100%. The modified problem formulation are presented as follows,

## Problem formulation

$$\min \quad TSC = B_r \times b + \sum_t VOLL_t \times AEL(t)$$

s.t. $AEL(t) = \begin{cases} D(t), & \text{if } Q_B(t) + \int_t^{t+\Delta t} \left(P(t) - D(t)\right) \times e \, dt < B_{\min} \\ 0, & \text{otherwise} \end{cases}$ (1)

$$Q_B(t + \Delta t) = \begin{cases} \min\left\{Q_B(t) + P(t) \times e(t), B_r\right\}, & \text{if } Q_B(t) + \int_t^{t+\Delta t} \left(P(t) - D(t)\right) \times e \, dt < B_{\min} \\ Q_B(t) + \int_t^{t+\Delta t} \left(P(t) - D(t)\right) \times e(t) \, dt, & \text{if } B_{\min} \leq Q_B(t) + \int_t^{t+\Delta t} \left(P(t) - D(t)\right) \times e \, dt \leq B_r \\ B_r, & \text{if } Q_B(t) + \int_t^{t+\Delta t} \left(P(t) - D(t)\right) \times e \, dt > B_r \end{cases}$$ (2)

$$P(t) = \eta \times I(t) \times A, \quad \forall \, t$$ (3)

$$B_{\min} = B_r \times \left(1 - DoD\right)$$ (4)

$$LOLP = \sum_{t \in DNS} \Delta t \Big/ T, \quad DNS = \left\{t; \, Q_B(t) + \int_t^{t+\Delta t} \left(P(t) - D(t)\right) \times e \, dt < B_{\min}\right\}$$ (5)

$$CCP = \Pr\left\{LOLP \leq \beta\right\} \geq 1 - \alpha$$ (6)

when $t = 0$,

$$\Pr\left\{Q_B(t = 0)\right\} = \begin{cases} 0, & Q_B(0) < B_{\min} \\ \dfrac{1}{B_r - B_{\min}}, & 0 \leq Q_B(0) \leq B_r \\ 0, & Q_B(0) > B_r \end{cases}$$ (7)

$$0 \leq \alpha, \beta \leq 1, \quad B_{\min} \geq 0, \quad B_r \geq 0$$

The objective function includes two terms, $B_r \times b$ and $\sum_t VOLL_t \times AEL(t)$. The first term represents the battery investment cost while the second term denotes the penalty cost for unsatisfied load demand. The optimal decision made considering the trade-off between these two terms is one of the main cores of the current research. It can be easily proven that there is a positive correlation between battery capacity and the first term whereas a negative correlation between battery capacity and the second term. It is worth mentioning that the reliability metric used is chance constraint probability (*CCP*) which is presented in eq. 6 along with its minimum level, 1 - $\alpha$. In this context, $\alpha$ and $\beta$ represent the reliability strictness of the system modeler. $Q_B(t)$ is the amount of energy stored in battery (kWh) at time $t$, while the charging state of battery, $Q_B(t)$, at the beginning of each grid outage is assumed to follow uniform distribution. More details of the formulation can be found in [12].

## 3. Case study and results

In this section, the proposed simulation-based optimization method is tested on each of the four battery types mentioned before separately and then the results are compared against each other. The demand and solar irradiation data used are referring to a hospital facility in Orlando, FL [13, 14]. *VOLL* is selected to be in the maximum level of the range provided in [15], due to the critical nature of the considered facility. In this paper, PV array conversion efficiency is set to be 16% [16]. CAIDI and SAIFI are 7.83 and 0.84 for Orlando, FL [17] and the *LOLP* constraint is set to be 10%. The solar array size is equal to 20,000 m² and the battery capacities are ranging from 200 kWh to 20,000 kWh. Finally, the important battery properties used in our model are projections for year 2030 based on [18] and are presented in Table 1. The analysis is conducted for a planning time horizon of 20 years and future costs are discounted at a rate of 4% to determine the present value for the entire planning horizon.

**Table 1 Approximate central estimates for year 2030 of cost, efficiency and DoD for four battery types [18]**

| Parameter<br>Battery Type | Cost ($/kWh) | Efficiency (%) | DoD (%) |
|---|---|---|---|
| Lead-Acid | 75 | 86 | 55 |
| Sodium Sulphur | 165 | 86 | 100 |
| Vanadium Redox | 120 | 78 | 100 |
| Lithium-ion | 224 | 97 | 90 |

It can be seen from Table 1 that the lead-acid battery is expected to remain the least expensive battery. However, the cost gap is going to close between this type and the other types of batteries, especially with the vanadium redox flow battery based on [18]. Moreover, it would be interesting to explore whether the superiority of certain battery types in terms of *DoD* (sodium sulphur, vanadium redox) can overcome their pricing barriers. In Fig. 1, results can be seen for all types of batteries in a dual axis format for *TSC* and *CCP* with predetermined *LOLP* constraint of 10%:

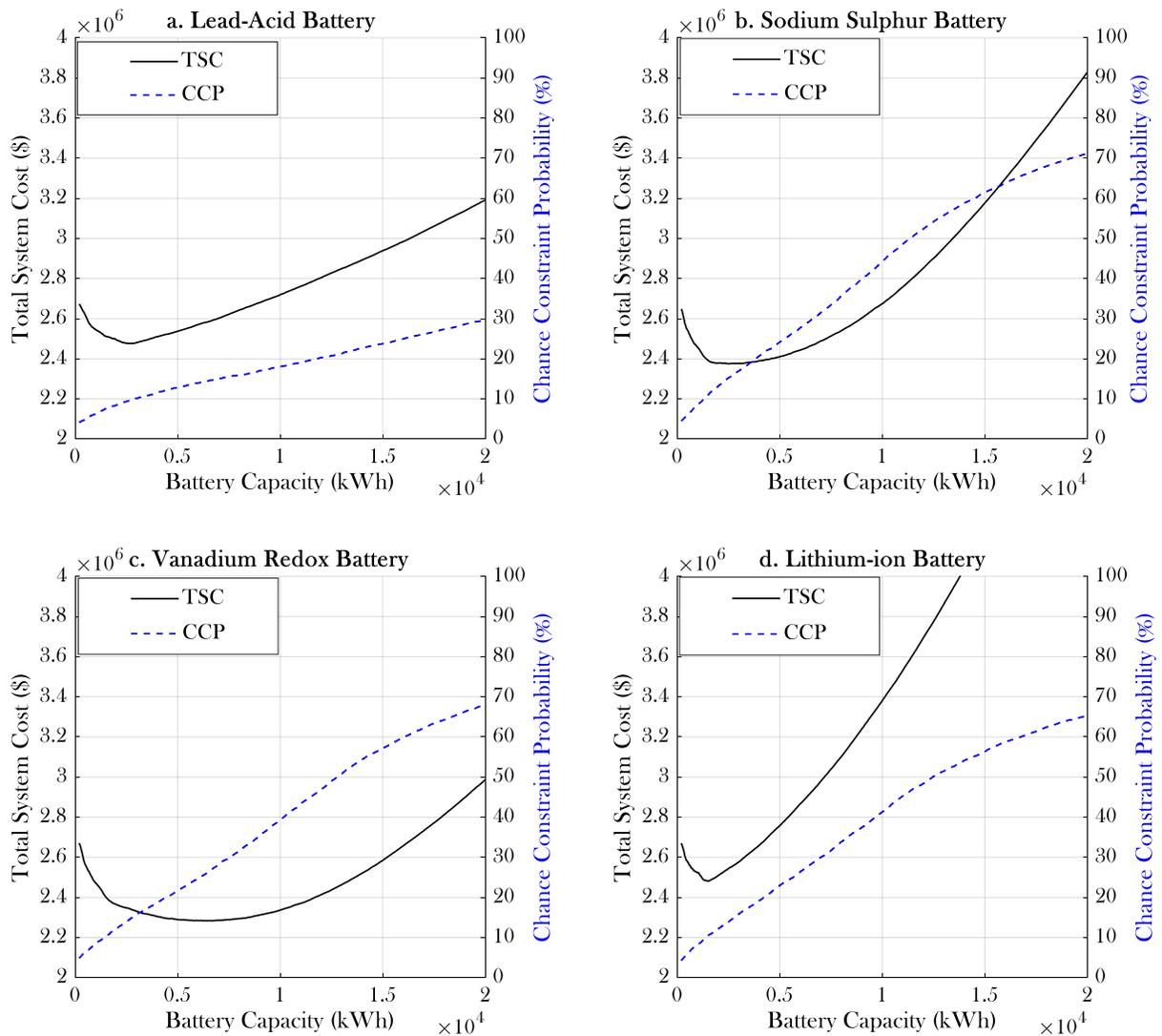

**Fig. 1. TSC and CCP vs Battery Capacity for all four battery types.**

There are some very important and intuitive conclusions that can be drawn from Fig. 1, and this can be done through individual or pairwise observations. Firstly, it can be shown that lead-acid battery achieved the worst performance in terms of *CCP* and this is almost exclusively due to its extremely low *DoD*. Lithium-ion was the most expensive battery and this can be easily justified from the highest comparative *TSC*. On the contrary, it should be noticed that vanadium redox battery is a storage type with great potential because *TSC* stays low as battery capacity of this type changes in the considered range with a satisfactory level of *CCP*. Lastly, sodium sulphur battery type achieved the highest *CCP*, although the differences are small when compared with that achieved by vanadium redox and lithium-ion battery types. While the corresponding *TSC* is between that of vanadium redox and lithium-ion battery types.

An interesting comparison that should also be examined carefully is between sodium sulphur battery and vanadium redox battery. It should be reminded here that the former type had a higher efficiency than that of the latter one, as well as a higher cost. However, their differences in *TSC* are much more significant than their differences in *CCP*. It demonstrates that vanadium redox battery is more cost-effective than sodium sulphur battery. Another worthwhile comparison is between vanadium redox battery and lithium-ion battery. The 10% difference in the *DoD* of these two batteries was able to offset their higher difference of 19% in terms of efficiency, resulting in a very similar *CCP* assessment, so cost becomes the dominant factor for choosing between these two batteries. This provides convincing evidence of the importance of *DoD* in selecting the appropriate battery type for photovoltaic and battery systems from resilience and economic aspects.

## 4. Conclusions

Four different battery types, based on their forecasted properties for the year 2030, were compared as a part of photovoltaic and battery systems in terms system resilience and economic viability. For this purpose, an existing simulation-based optimization method was improved and used in a case study with real demand and irradiation data in this paper. Based on the simulation results, a general conclusion that can be drawn is the high importance of the three selected battery properties, i.e. cost, efficiency and *DoD* regarding the performance of photovoltaic and battery systems. More specifically, *DoD* and cost play vital roles in the optimal configuration of these systems. This finding, along with the anticipated decreasing cost of flow batteries, could be the reason behind the increasing attention that is given to flow batteries, like vanadium redox battery. As a direction for future research, the importance for investigation of more battery types should definitely be mentioned, as well as the need for developing more advanced and analytical forecasting techniques specifically focused on battery properties.